\documentclass[prb,aps,espf, twocolumn]{revtex4}

\usepackage{amsmath}
\usepackage{graphicx}
\usepackage{epstopdf}
\begin{document}

\textbf{Rapid Research Letters (Accepted, 2019)}
\title{CrAs monolayer: Low buckled two-dimensional half-metal ferromagnet  }

\author{Gul Rahman}\email{gulrahman@qau.edu.pk}
\affiliation{Department of Physics,
Quaid-i-Azam University, Islamabad 45320, Pakistan}

\author{Zakir Jahangirli}
\affiliation{Institute of Physics of ANAS, G. Javid ave. 131, AZ1143, Baku, Azerbaijan}
\affiliation{Azerbaijan Tecnical University, G. Javid ave. 25, AZ1073, Baku, Azerbaijan}

\begin{abstract}
\textit{Ab-initio} calculations based on density functional theory (DFT) are performed to study the structural, electronic, and magnetic properties of two-dimensional (2D) free-standing honeycomb  CrAs. We show that CrAs has low buckled stable structure. Magnetic CrAs has larger buckling than non-magnetic CrAs.  
2D-CrAs is a ferromagnetic semiconductor for lattice constant $a \leq 3.71$\AA, and above this lattice constant CrAs is a half-metal ferromagnet.
2D-CrAs is shown to be half-metal ferromagnetic with magnetic moment of 3.0$\mu_{\rm{B}}$ per unit cell, at equilibrium structure.
The $d_{z}^{2}$ orbital of $e_{g}$ band is completely empty in the spin-down state whereas it is almost occupied in the spin-up state, and the magnetic moment in the $e_{g}$ band is mainly dominated by the $d_{z}^{2}$ orbital of Cr.
The $d_{zx}/d_{zy}$ and $d_{xy}$ orbitals of $t_{2g}$ band are partially occupied in the spin-up state and behaves as metal whereas they are insulator in the spin-down state.
Phonon calculations confirm the thermodynamic stability of 2D-CrAs.  
The ferromagnetic (FM) and antiferromagnetic (AFM) interaction between the Cr atoms reveal that the FM state is more stable than the AFM state of 2D-CrAs.

\end{abstract}


\maketitle

Much attention has been devoted to discover new two-dimensional (2D) materials due to their exceptional properties such as high electrical conductivity, mechanical strength, band tunability, after the discovery of graphene.\cite{Andrie,g1,g2,g3}
New 2D elemental materials with a band-gap such as silicene, germanene, stanene, phosphorene, arsenene, etc.\cite{n1,n2,n3,gr2014,p1,p2,p3,kamal2015} have nevertheless been discovered in the last decade. Most of these 2D materials are non-magnetic semiconductors. To utilize these 2D materials in spintronic devices, either intrinsic defects or magnetic impurity atoms are used to induce magnetism in 2D materials.\cite{RF} In the recent past we explicitly focused on developing magnetism in non-magnetic 2D materials.\cite{GR,GR2} However, it costs energy to induce magnetism in non-magnetic 2D materials. Hence, it is quite natural to search for intrinsic ferromagnetic (FM) 2D materials
for possible usages in carrier injection,
detection, sensors, magnetic storage, and emergent heterostructure devices.
The quest for 2D FM materials\cite{GR2017} is very young and efforts are under way to discover new intrinsic FM materials that can replace the conventional bulk magnetic materials.

Recently, the successful fabrications of 2D
FM  semiconductors, e.g., CrI$_{3}$\cite{ref1} and
Cr$_{2}$Ge$_{2}$Te$_{6}$,\cite{ref2} magnetism in 2D materials got particular attention due to their potential for nanoscale device
applications. Experimentally, Cr$_{2}$Ge$_{2}$Te$_{6}$ was achieved through exfoliation, and CrI$_{3}$ was obtained
and proved to be an Ising ferromagnet with Curie temperature
T$_{c}$ of 45 K. 
First-principles calculations also show that monolayers of  CrSnTe$_{3}$,  CrSiTe$_{3}$, and CrGeTe$_{3}$ are FM semiconductors.\cite{prb2015} 
The magnetic moment carried by the Cr$^{3+}$ ions, which are arranged in a honeycomb
lattice structure and octahedrally coordinated by nonmagnetic I- ions, is aligned in the
out-of-plane direction by anisotropic exchange interaction mediated by the I-ions.\cite{ref20, ref21, ref22} Here we propose 2D-CrAs which also shares the honeycomb like arrangement of Cr atoms, hence it is expected that monolayer CrAs can be another 2D FM material.
Bulk CrAs got particular attention in 2000 when Akinaga \textit{et al.}\cite{Akinaga}  announced the successful synthesis of metastable Zincblende(ZB)-type CrAs on GaAs substrate using molecular beam epitaxy (MBE), and they found FM
behavior at room temperature. Bulk NiAs-type and MnP-type CrAs are found to be antiferrimagnetic (AFM) metals,\cite{niasCrAs,peter} whereas ZB-type CrAs is half-metal FM. The main purpose of the present work is to study the electronic and magnetic structures of monolayer of honeycomb like CrAs (see Fig.S1), and to investigate whether CrAs prefers planar, low buckled, high buckled, FM, AFM, metallic, semiconductor, metal, or half-metallic structure.

\textit{Ab-initio} calculations based on density functional theory (DFT) are performed with the plane-wave and pseudopotential method implemented in the Quantum Espresso package.\cite{26GiannozziSBaroni} The exchange and correlation energy and potential were calculated with the Perdew-Burke-Ernzerhof (PBE) parametrization \cite{27JPPerdew1996} of the generalized gradient approximation (GGA). The ultrasoft pseudopotentials were parameterized with the recipe of Rappe, Rabe, Kaxiras and Joannopoulos.\citep{28RRKJ}  The electron wave function was expanded in a plane wave basis set cut-off of 50 Ry.  A dense $20\times 20\times 1$ Monckhorst-Pack grid \cite{28Monkhorst} was used for the $k$-points, which gave a fine reciprocal-space grid and hence a rather high accuracy. A vacuum slab of 15 \AA\, was used in the direction normal to the  plane of arsenene to ensure the absence of interlayer interactions in that direction. The convergence of all computational parameters was checked carefully. {Phonon dispersion calculations were performed based on the DFPT\cite{dfpt1} using the ABINIT code.\cite{abinit}}


Before we present our work on monolayer CrAs, attention will be given to the bulk CrAs. The structural and magnetic properties of bulk CrAs are studied using NiAs-type (hexagonal), MnP-type CrAs, and ZB-type  CrAs.  All the calculations were carried out in non-magnetic (NM, non-spin polarized) and magnetic (M, spin polarized) states.
From the DFT total energy calculations, the cohesive energy $E_{c}$ is evaluated as $E_{c}$=E(CrAs)-[E(Cr)+E(As)], where E(CrAs), E(Cr), and E(As) is the total energy of CrAs in the respective phase, atomic Cr, and  atomic As, respectively.
Fig.S2 summarizes the cohesive energy vs lattice constant of CrAs.  Our calculated lattice volume per formula unit (f.u.) of MnP-, ZB- and NAs-type CrAs in M state is 33.4\AA$^{3}$, 46.6\AA$^{3}$ and 33.38\AA$^{3}$, respectively. These values are comparable with the previous values.\cite{peter}. The calculated cohesive energy/f.u. of ZB- and NAs-type CrAs in the M state is -6.4 eV and -6.8 eV, respectively. It shows that bulk NiAs-type CrAs is more stable than ZB-type CrA. The calculated magnetic moment per unit cell of ZB- and NAs-type CrAs is 3.0$\mu_{\rm{B}}$ and 4.85$\mu_{\rm{B}}$, respectively. The electronic structures revealed that ZB-type CrAs is half-metal, whereas NiAs-type CrAs is a ferromagnetic metal. 
These values also agree with the previous theoretical values.\cite{peter}

\begin{figure}[!]
   \centering
   \includegraphics[width=.3\textwidth]{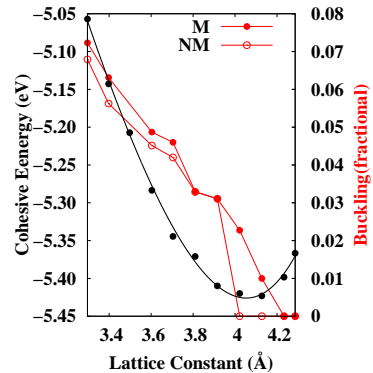}
  \caption{Calculated cohesive energy  (left) and optimized buckling (right) vs lattice constant  of 2D-CrAs. The black solid line/circle shows the cohesive energy in the M state, whereas the red solid (open) circle shows the buckling in M (NM) state.}
        \label{Buckling}
\end{figure}

Mostly monolayers prefer either buckled or puckered structure except graphene which adopts the planar geometry. Thus it is very essential to investigate whether our proposed material adopts the planar, low buckled, or high buckled structure.
A buckled structure is always expected to sustain larger tensile strain than a planar structure. 
Hence, it is very important to find the optimized structure of 2D-CrAs, and for this purpose we carried out extensive total energy calculations using different lattice constant $a$ and buckling height $\delta$.   
Fig.{S3} shows the calculated cohesive energy per unit cell vs buckling of 2D-CrAs in the NM and M states. It is clear to see that the M state always has lower energy than the NM state, hence, it is legitimate to call 2D-CrAs as magnetic 2D-CrAs. The smaller lattice constant prefers highly buckled structure, whereas the larger lattice constant keeps CrAs in a low buckled state. As the lattice constant is increased, the buckling is washed out which is consistent with our recent work on arsenene.\cite{GR2019}
From these calculations it is possible then to infer that each lattice constant has its own buckling parameter, which will have a strong influence on the physical properties of each case.\cite{GR2019}
Using the data of Fig.{S3}, the buckling parameter for each lattice constant and the optimized lattice constant are summarized in Fig.\ref{Buckling}. 

Fig.~\ref{Buckling} shows the optimized buckling calculated for each lattice constant in M and NM case. The buckling height decreases with lattice constant in either case. Magnetic CrAs always have larger buckling as compared with NM. The larger value of buckling in the M state is due to magnetostriction in CrAs. CrAs prefers planar structure in N (M) state for lattice constant $\geq 4.0$ (4.2)\AA. 
Fig.\ref{Buckling} also shows the cohesive energy of 2D-CrAs in the M state, and one can see that the optimized $a$ is 4.05\AA,~ and buckling is 0.49\AA.{Note that the value of buckling in 2D-CrAs is different from elemental 2D-materials, i.e., graphene, silicene, etc.\cite{sili} } 
The cohesive energy of 2D-CrAs is also shown in Fig.{S1} for comparison.
The calculated cohesive energy is evaluated to be -5.4 eV, which is (1.0 eV) 1.4 eV larger than the (ZB) NiAs-type CrAs. We must stress that ZB-CrAs is also a metastable phase of CrAs, but it has been successfully achieved experimentally over a suitable substrate.\cite{Akinaga} Thus we believe that 2D-CrAs could also be achieved.
The calculated $a$ and $\delta$ of CrAs can be compared with arsenene which are
3.61\AA~ and 1.39\AA, respectively.\cite{GR2019}
So, it can be inferred that Cr doping in arsenene will increase the lattice constant and decrease the buckling, which is in agreement with our analysis of Fig.{S3}. 
We must note that the optimized lattice constant in the M case lies between 4.0 and 4.2\AA,~(see Fig.\ref{Buckling}) which implies that 2D-CrAs has low buckled structure. 
It is also essential to mention that in NiAs-type CrAs the buckling height is about 1.34\AA. Therefore, reducing the dimensionality (making monolayer) of a material is also reducing the buckling height which will also change the hybridization between Cr and As atoms.

\begin{figure}[!]
   \centering
   \includegraphics[width=0.3\textwidth]{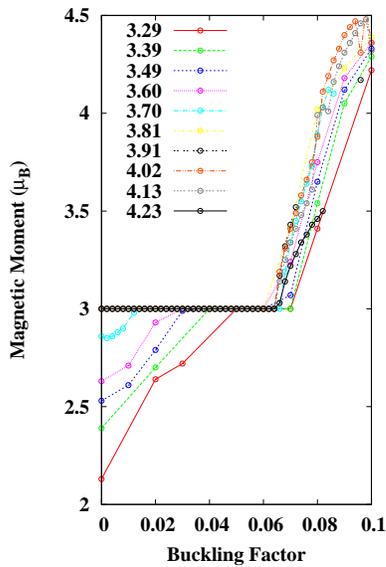}
 \caption{Calculated magnetic moment (in $\mu_{\rm{B}}$) per unit cell vs buckling of 2D-CrAs for different lattice constants (in \AA).}
        \label{MM}
\end{figure}
Fig.~\ref{MM} shows the calculated magnetic moment per unit cell as a function of buckling for different lattice constants. For the planar CrAs, the magnetic moment increases as the lattice constant increases and attains a maximum value of 3.0$\mu_{\rm{B}}$ when $a > 3.70$\AA. The magnetic moment also increases with buckling and attains maximum value 3.0$\mu_{\rm{B}}$, the buckling at which the magnetic moment saturates shifts to lower buckling as the lattice constant is increased. For $a > 3.70$\AA~ the magnetic moment remains 3.0$\mu_{\rm{B}}$ until the buckling is $\le 0.08$, and beyond this buckling the magnetic moment is further increased to 4.5$\mu_{\rm{B}}$. As the lattice is increased the buckling is reduced and the low buckled structure always attains the integer value of magnetic moment---a prerequisite for half-metal ferromagnet.
{Note that the calculated magnetic moment 3.0$\mu_{\rm{B}}$ of 2D-CrAs in the ground state indicates that the oxidation state of Cr is $+3$. However, the increase in the magnetic moment with buckling mainly happens due to charge transfer\cite{stefano} and this charge transfer may change the oxidation state of Cr in CrAs. We must also note that the oxidation  state of Cr is different in different compounds.\cite{Noce} }

It will be interesting to investigate the electronic structure of 2D-CrAs in the NM and M states to see how Cr and As bond with each other and how magnetic moment is formed. Fig.{S4} shows the calculated electronic density of states (DOS) at optimized $a$ and $\delta$ in the NM state. For all lattice constants, 2D-CrAs has large DOS at the Fermi level, i.e., CrAs shows metallic behavior in the NM state. The $d$-orbital is further decomposed into $t_{2g}$ and $e_{g}$ states. The  $t_{2g}$ and $e_{g}$ electrons stay at the same energy at smaller lattice constant. However, at larger $a$ (7.6 Bohr), an isolated sharp peak at 2 eV below the Fermi energy can be seen and this sharp peak is mainly contributed by the $t_{2g}$ electrons of Cr. Due to the same orbital symmetry of $t_{2g}$ and $p$ electrons of As, strong bonding can be seen. The metallic character in the NM state is mainly contributed by the  $t_{2g}$ electrons of Cr. Nevertheless, the large DOS at the Fermi energy is an indication of magnetic instability.

%

Fig.{S5} shows the calculated total DOS and projected DOS (PDOS) in the spin-polarized case. One can clearly see that 2D-CrAs is a magnetic semiconductor for $a \leq 7.0$ (Bohr), and above this lattice constant CrAs is true half-metal ferromagnet. Thus we can claim that compressive strain drives 2D-CrAs into semiconducting state, whereas tensile strain brings 2D-CrAs into half-metallic state. The half-metallic character is consistent with the integer magnetic moment of CrAs. One sees that $t_{2g}$ and $e_{g}$ bands are partially occupied in the spin-up states, whereas the $e_{g}$ band is completely empty in the spin-down states. Therefore, it can be concluded from this figure that strain can play a major role to tune the electronic structure of 2D-CrAs. The large bandgap in the spin-down states suggest that 2D-CrAs can be a possible candidate for spintronics where 100\% spin-polarization is needed.  

\begin{figure}[!]
   \centering
   \includegraphics[width=0.3\textwidth]{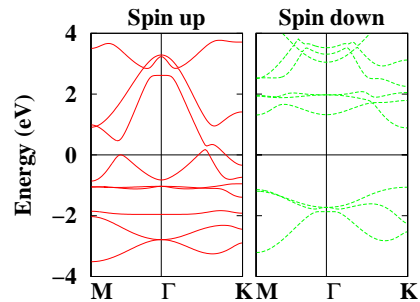}
  \caption{Calculated electronic band structure of 2D-CrAs. The horizontal line shows the Fermi energy, which is set to zero eV.}
        \label{band-Equal}
\end{figure}

In Fig.~\ref{band-Equal}
we present the corresponding spin-dependent energy bands of 2D-CrAs at optimized $a$ and $\delta$ along high-symmetry directions in the Brillouin zone. There
is one band (M-$\Gamma$-K) in the spin-up state which is crossing the Fermi energy making it metallic. This band has a hole like character. On the other hand, one can see two bands are crossing the Fermi energy near K-point (K-$\Gamma$). The crossing band at K-point has electronic like behavior. In the spin-up state a non-bonding band, contributed by Cr $d$ orbitals, around 1 eV below the Fermi energy can be seen. The spin-down band does not show any band at the Fermi energy and it behaves as a semiconductor. There is about
2.40 eV bandgap at the M point and 1.90 eV bandgap at the K-point in the spin-down state. Hence, 2D-CrAs has a half-metallic band structure and such a band structure can play an essential role in the transport properties of spintronic nano devices. {Careful analysis of the band structure of 2D-CrAs shows that the bandwidth of the 2D-CrAs for $p+d$ is about 7 eV which is lesser than 12-13 eV of the 3D-CrAs.\cite{philos} Hence, 2D-CrAs seems a strongly correlated system differently from 3D-CrAs.\cite{gr}
} {
We must stress that electrons close to the Fermi energy are very important from transport point of view, and it is very essential to see the Fermi surface of 2D-CrAs.  The Fermi surfaces of 2D-CrAs are calculated in the NM and M states (see Fig. S6). As the NM 2D-CrAs shows metallic behavior and the calculated Fermi surface shows two concentric cylinders centered at the BZ of CrAs.  Once can also see small six electron pockets. The spin-up Fermi surface composed of two bands that cross the Fermi energy and it has hole like nature, whereas one band has electronic like nature.}
\begin{figure}[!]
   \centering
    \includegraphics[width=0.3\textwidth]{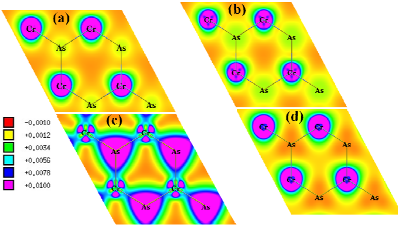}
  \caption{Spin Density contours of FM 2D-CrAs (a) VB Spin up (b) CB Spin Up , (c) VB Spin Down (d) CB Spin Down, inset shows the scale }
        \label{SpinDens}
\end{figure}

At the optimized $a$ and $\delta$, the total magnetic moment is calculated to be 3.0$\mu_{\rm{B}}$ per unit cell. Our further analysis of Lowdin charges shows that the magnetic moment is mainly contributed by $t_{2g}$ and $e_{g}$ orbitals of Cr that give about 2.07~$\mu_{\rm{B}}$ and 1.47~$\mu_{\rm{B}}$, respectively. On the other hand, the $p$ electrons of As also has a small induced magnetic moment of -0.70~$\mu_{\rm{B}}$. 
Fig.\ref{SpinDens} further shows the calculated spin density in the magnetic state in the spin-up and spin-down states near the conduction and valance band edges. In spin-up state the spin density is mostly localized around the Cr atoms and minor contribution from the As atoms, whereas the contribution of As atoms is slightly increased in the conduction band. The spin density around the As atoms is spherically distributed which implies a covalent bonding with Cr atoms. In the spin-down case, the spin density is mostly localised around the As atoms near the valence band edge and the spin density around the As atoms is not spherically distributed but rather pointing towards the Cr atoms. Close inspection of Fig.~\ref{SpinDens} shows that the Cr (As) spin density has $d_{xz}/d_{yz}$ ($p_{z}$)-like character and no orbitals ($t_{2g}$--$p$) repulsion can be seen. Such a polarized density between the Cr and As atoms brings them in the bonding state. On the other hand, no spin density around the As atoms in the conduction band can be seen in the spin-down state which implies that the As orbitals are completely occupied by electrons. The spin density is also mostly localized around the Cr atoms.
\begin{figure}[!]
   \centering
   \includegraphics[width=0.25\textwidth]{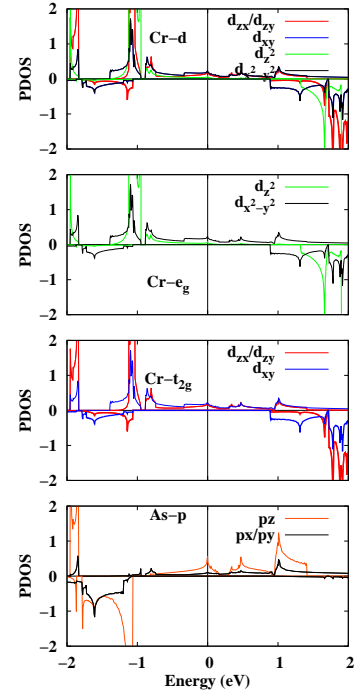}
   \caption{Partial Density of State of 2D-CrAs. The vertical line shows the Fermi energy ,which is set to zero eV. The upper (middle) panel shows the $e_{g}$ ($t_{2g}$) orbitals of Cr, whereas the lower panel shows the $p$ orbitals of As atoms. The top panel shows the combined $e_{g}$  and $t_{2g}$ orbitals of Cr. Note that the positive (negative) PDOS on the vertical axis shows spin-up (spin-down) DOS.  }
        \label{DOS-Equal}
\end{figure}

To see the different contribution of bands and the orbital origin of magnetism, in Fig.\ref{DOS-Equal} the $t_{2g}$ and $e_{g}$ orbitals of Cr and $p$ orbital of As are further decomposed. The $d_{z}^{2}$ orbital is completely empty in the spin-down state whereas it is almost occupied in the spin-up state. The $d_{{x}^{2}-{y}^{2}}$ electrons of the $e_{g}$ band in the spin-up state is itinerant, whereas they are localized in the spin-down state. Thus it is elusive that the magnetic moment in the $e_{g}$ band is mainly dominated by the $d_{z}^{2}$ orbitals of Cr. The $d_{zx}/d_{zy}$ and $d_{xy}$ orbitals of the $t_{2g}$ band are partially occupied in the spin-up state and behave as metal whereas they are insulator in the spin-down state. The $p$ orbital of As is also effected by the crystal field of the hexagonal symmetry. These are the hybridized $p_{z}$ electrons of As with Cr-$d$ that contribute to metallicity in the spin-up state, whereas the spin-down band is completely occupied. Hence, due $p$-$d$ hybridization all electrons are partially occupied in the spin-up band of As and the spin-down band remains completely filled.
Within the double-exchange model, the conducting electrons in the hybridized
bands can play a vital role in establishing ferromagnetism
in 2D-CrAs. {Fig.\ref{DOS-Equal} also shows the combined PDOS of $t_{2g}$ and $e_{g}$  and one can clearly see that the $d_{{x}^{2}-{y}^{2}}$ and $d_{xy}$  orbitals strongly hybridize because they are symmetrically equivalent. }

\begin{figure}[!]
   \centering
   \includegraphics[width=0.3\textwidth]{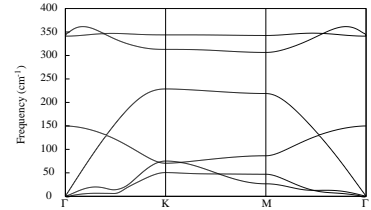}
   \caption{Calculated phonon dispersions of 2D-CrAs.  }
        \label{phon}
\end{figure}

{Before discussing the FM and AFM interaction, it is necessary to address the thermodynamic stability of 2D-CrAs, which is investigated using phonon.  The phonon dispersions along selected high symmetry directions in the BZ are shown in Fig.\ref{phon}. As can be seen from Fig.\ref{phon}, the phonon spectra reveal positive frequencies along all directions, which indicates that 2D-CrAs is dynamically stable. Further experimental work would be needed to synthesize monolayer of CrAs.}
{To explore potential applications of 2D-CrAs in magnetic storage, it is essential to calculate its magnetic anisotropy energy (MAE). Our preliminary results show that 2D-CrAs exhibits a large MAE of 1.20 meV per Cr atom.}
Finally, using the optimized $a$, we considered a $2\times 2\times 1$ supercell to consider FM and AFM coupling between the Cr atoms. Interestingly we found that the FM state is more stable than the AFM state by 28 meV/f.u. (there are 4 f.u. in $2\times 2\times 1$ ) of CrAs. We also checked the FM stability of planar CrAs against the AFM state, and we observed that the planar CrAs also prefers the FM state by 25 meV/f.u. against the AFM state. We must note that the bulk CrAs has AFM ground state,\cite{niasCrAs,peter} but the magnetic interaction  strongly depends on the Cr-Cr separation.\cite{prx}
{To estimate the Curie temperature $T_{c}$, we map the energetics of 2D-CrAs onto an Ising model Hamiltonian on a triangular lattice with nearest neighbor interactions between the magnetic spins, and followed our previous procedure.\cite{GR2017} Our DFT estimated $T_{c}$ is about 150 (137) K for buckled (planar) 2D-CrAs, and therefore ferromagnetism
in monolayer of CrAs is expected.}
Thus we believe that 2D-CrAs can be a true half-metal ferromagnetic around room temperature.

To summarize, extensive \textit{ab-initio} calculations were carried out to probe the stable electronic and magnetic structures of 2D-CrAs.  2D-CrAs has low buckled  structure. The electronic structure of 2D-CrAs shows 100\% spin polarization with magnetic moment of 3.0~$\mu_{\rm{B}}$. FM 2D-CrAs is more stable than AFM one. The thermodynamic stability of 2D-CrAs is checked using phonon calculations. We believe that 2D-CrAs can be an other material for nanoelectronic spintronic devices.




\begin{thebibliography}{99}





\bibitem{Andrie} Novoselov, K. S., Geim, A. K., Morozov, S. V., Jiang, D. A., Zhang, Y., Dubonos, S. V., Grigorieva, I. V. \& Firsov, A.A. \textit{Science}, \textbf{306}, 666 (2004).
\bibitem{g1}Neto, A. C., Guinea, F., Peres, N. M., Novoselov, K. S. \& Geim, A. K. \textit{Rev. Mod. Phys.}, \textbf{81}, 109 (2009).
\bibitem{g2}Sahdan, M. F., \& Darma, Y. \textit{AIP Conference Proceedings}, \textbf{1589}, 253 (2014).
\bibitem{g3}Peng, Q., Dearden, A. K., Crean, J., Han, L., Liu, S., Wen, X. \& De, S. \textit{Nanotechnol. Sci. Appl.}, \textbf{7}, 1 (2014).

\bibitem{n1}Ezawa, M. \textit{J. Phys. Soc. Jpn.}, \textbf{84}, 121003 (2015).
\bibitem{n2}Acun, A. \textit{et al.} \textit{J. Phys.: Condens. Matter.}, \textbf{27}, 443002 (2015).
\bibitem{n3}Lu, P. \textit{et al.} \textit{Sci. Rep.}, \textbf{7}, 3912 (2017).

\bibitem{gr2014}Rahman, G. \textit{EPL}, \textbf{105}, 37012 (2014).

\bibitem{p1}Buscema, M., Groenendijk, D. J., Steele, G. A., Van Der Zant, H. S. \& Castellanos-Gomez, A. \textit{Nat. Commun.}, \textbf{5}, 4651 (2014).
\bibitem{p2}Liu, H., Neal, A. T., Zhu, Z., Luo, Z., Xu, X., Tománek, D. \& Ye, P. D. \textit{ACS Nano}, \textbf{8}, 4033 (2014).
\bibitem{p3}Xia, F., Wang, H. \& Jia, Y. \textit{Nat. Commun.}, \textbf{5}, 4458 (2014).
\bibitem{kamal2015}Kamal, C. \& Ezawa, M. \textit{Phys. Rev. B}, \textbf{91}, 085423 (2015).


\bibitem{RF} Zhang, Q.,Ren Z., Wu, N., Wang, W.,Gao, Y.,Zhang, Q., Shi, S., Zhuang, L., Sun, X., \& Fu, L. \textit{npj 2D Materials and Applications}, \textbf{2}, 22 (2018). 

\bibitem{GR} Rahman, A. U., Rahman, G., \& Garc\'{\i}a-Su\'arez, V. M. \textit{J. Magn. Magn. Mater.} \textbf{443}, 343 (2017).

\bibitem{GR2} Rahman, A. U., Rahman, G., Kratzer, P.
\textit{J. Phys.: Condens. Matter.}, \textbf{30}, 195805 (2018).

\bibitem{GR2017} Rahman, G.,Rahman, A. U., Kanwal, S., Kratzer, P.
\textit{EPL}, \textbf{119}, 57002 (2017).


\bibitem{ref1}Huang, B., Clark, G., Navarro-Moratalla, E., Klein, D.R., Cheng, R., Seyler, K.L., Zhong, D., Schmidgall, E., McGuire, M.A., Cobden, D.H. \& Yao, W. \textit{Nature}, \textbf{546}, 270 (2017).

\bibitem{ref2}Gong, C., Li, L., Li, Z., Ji, H., Stern, A., Xia, Y., Cao, T., Bao, W., Wang, C., Wang, Y. \& Qiu, Z.Q. \textit{Nature}, \textbf{546}, 265 (2017).






\bibitem{prb2015}Zhuang, H.L., Xie, Y., Kent, P. R. C. \& Ganesh, P. \textit{Phys. Rev. B}, \textbf{92}, 035407 (2015).

\bibitem{ref20} McGuire, M.A. \textit{Crystals}, \textbf{7}, 121 (2017).

\bibitem{ref21} McGuire, M.A., Dixit, H., Cooper, V.R. \& Sales, B.C. \textit{Chem. Mater.}, \textbf{27}, 612 (2015).

\bibitem{ref22} Lado, J. L., \& Fernández-Rossier, J. \textit{2D Mater.}, \textbf{4}, 035002 (2017).





\bibitem{Akinaga} Akinaga, H., Manago, T., \& Shirai, M. Jpn. J. Appl. Phys., Part
2 \textbf{39}, L1118 (2000).

\bibitem{niasCrAs} Selte, K, Kjekshus, A., Jamison,  W. E.,  Andresen,  A. F., \&  
Engebretsen, J. E. \textit{Acta Chem. Scand}. (1947-1999) \textbf{25}, 1703 (1971).


\bibitem{peter} Hashemifar, S. J., Kratzer, P., \& Scheffler, M. \textit{Phys. Rev. B}, \textbf{82}, 214417 (2010).














\bibitem{26GiannozziSBaroni}Giannozzi, P., Baroni, S., Bonini, N., Calandra, M., Car, R., Cavazzoni, C., Ceresoli, D., Chiarotti, G.L., Cococcioni, M., Dabo, I. \& Dal Corso, \textit{J. Phys. Condens. Matter}, \textbf{21}, 395502 (2009).



\bibitem{27JPPerdew1996}Perdew, J. P., Burke, K. \& Ernzerhof, M. \textit{Phys. Rev. Lett.}, \textbf{77}, 3865 (1996).


\bibitem{28RRKJ}Rappe, A. M., Rabe, K. M., Kaxiras, E. \& Joannopoulos, J. D. \textit{Phys. Rev. B}, \textbf{41}, 1227 (1990).







\bibitem{28Monkhorst}Monkhorst, H. J. \& Pack, J. D. \textit{Phys. Rev. B}, \textbf{13}, 5188 (1976).

\bibitem{dfpt1}
Gianozzi, P., de Gironcoli, S., Pavone, P.\&  Baroni, S.  \textit{Phys. Rev. B} \textbf{43}, 7231 (1991).

\bibitem{abinit}
Gonze, X., Beuken, J. M.,  Caracas, R., Detraux, F.,  Fuchs, M.,  Rignanese, M., Sindic, L., Verstraete, M., Zerah, G. \& Jallet, F.  \textit{Comput. Mater. Sci.} \textbf{25}, 478 (2002).





\bibitem{GR2019}Rahman, G., Mahmood, A., \& García-Suárez, V. M. \textit{Sci. Rep.}, \textbf{9}, 7966 (2019).

\bibitem{sili}Balendhran, S., Walia, S., Nili, H., Sriram, S. \& Bhaskaran, M. Elemental analogues of graphene: silicene, germanene, stanene, and phosphorene. \textit{Small} \textbf{11}, 640 (2015).



\bibitem{stefano} Sanvito, S.,\& Hill, N. A. \textit{Phys. Rev. B}, \textbf{62}, 15553 (2000). 

\bibitem{Noce} Autieri, C., Cuono, G., Forte, F., Noce, C.  \textit{J. Phys.: Condens. Matt.} \textbf{29}, 224004 (2017).

\bibitem{philos} Autieria, C., \&  Noceb, C. \textit{Philosophical magazine}, \textbf{97} No 34, 3276 (2017).


\bibitem{gr}Further work would be needed to see the correlation between the bandwidth W and Coulomb potential U, i.e., U/W under stress/buckling. 



\bibitem{prx}Matsuda\textit{et.al.,}\textit{Phys. Rev. X}, \textbf{8},
031017 (2018).
\end{thebibliography}
\end{document}